\documentclass[aps,prl,twocolumn,superscriptaddress]{revtex4-1}
\usepackage{graphicx}
\usepackage{graphics}
\usepackage{amsmath}
\usepackage{natbib}

\usepackage{hyperref}

\begin{document}

\title{Electron-Hole Asymmetric Chiral Breakdown of Reentrant Quantum Hall States}

\author{A. V. Rossokhaty}
	\affiliation{Stewart Blusson Quantum Matter Institute, University of British Columbia, Vancouver, British Columbia, V6T1Z4, Canada}
	\affiliation{Department of Physics and Astronomy, University of British Columbia, Vancouver, British Columbia, V6T1Z4, Canada}
\author{Y. Baum}
    \affiliation{Department of Condensed Matter Physics, Weizmann Institute of Science, Rehovot 76100, Israel}
\author{J. A. Folk}
\email{jfolk@physics.ubc.ca}
	\affiliation{Stewart Blusson Quantum Matter Institute, University of British Columbia, Vancouver, British Columbia, V6T1Z4, Canada}
	\affiliation{Department of Physics and Astronomy, University of British Columbia, Vancouver, British Columbia, V6T1Z4, Canada}
\author{J. D. Watson}
	\affiliation{Department of Physics and Astronomy, and Station Q Purdue, Purdue University, West Lafayette, Indiana, USA}
	\affiliation{Birck Nanotechnology Center, Purdue University, West Lafayette, Indiana, USA}
\author{G. C. Gardner}
	\affiliation{Birck Nanotechnology Center, Purdue University, West Lafayette, Indiana, USA}
    	\affiliation{School of Materials Engineering, Purdue University, West Lafayette, Indiana, USA}
\author{M. J. Manfra}
	\affiliation{Department of Physics and Astronomy, and Station Q Purdue, Purdue University, West Lafayette, Indiana, USA}
	\affiliation{Birck Nanotechnology Center, Purdue University, West Lafayette, Indiana, USA}
	\affiliation{School of Electrical and Computer Engineering,  Purdue University, West Lafayette, Indiana, USA}
    	\affiliation{School of Materials Engineering, Purdue University, West Lafayette, Indiana, USA}

\date{\today}

\newcommand{\rxx}{{R}_{xx}}
\newcommand{\rxy}{{R}_{xy}}
\newcommand{\rdp}{{R}_{D}^+}
\newcommand{\rdm}{{R}_{D}^-}

\begin{abstract}
Reentrant integer quantum Hall (RIQH) states are believed to be correlated electron solid phases, though their microscopic description remains unclear.  As bias current increases, longitudinal and Hall resistivities measured for these states exhibit multiple sharp breakdown transitions, a signature unique to RIQH states.  A comparison of RIQH breakdown characteristics at multiple voltage probes indicates that these  signatures can be ascribed to a phase boundary between broken-down and unbroken regions, spreading chirally from source and drain contacts as a function of bias current and passing voltage probes one by one.  The chiral sense of the spreading is not set by the chirality of the edge state itself, instead depending on electron- or hole-like character of the RIQH state.
\end{abstract}


\maketitle
A variety of exotic electronic states emerge in high mobility 2D electron gases (2DEGs) at very low temperature, and in a large out-of-plane magnetic field. The most robust are the integer quantum Hall states, described by discrete and highly degenerate Landau levels. When the uppermost Landau level is partially filled, electrons in that level may reassemble into a fractional quantum Hall (FQH) liquid
\cite{laughlin, halperin, fqhe} or condense into charge-ordered states, from Wigner crystals to nematic stripe phases\cite{du, lilly, pan92,koulakovprl,koulakovprb, eisenstein02, xia2pk, csathy05, goerbig, shibata}.  Such charge ordered states, or electron solids, are observed primarily above filling factor $\nu=2$, where Coulomb effects are  strong in comparison to magnetic energy scales.  They are believed to be collective in nature\cite{csathy}, prone to thermodynamic phase transitions like melting or freezing of a liquid.

Numerical simulations of electron solids indicate alternating regions of neighboring integer filling factors with dimensions on the order of the magnetic length\cite{Fradkin:1999bl,Spivak:2006kf}. When the last Landau level is less than half-filled, the electron solid takes the form of ``bubbles" of higher electron density in a sea of lower density (an electron-like phase).  Above half filling, the bubbles are of lower electron density giving a hole-like phase. Insulating bubble phases lead to ``reentrant" transitions of the Hall resistivity up or down to the nearest integer quantum Hall plateau, giving rise to the term ``reentrant integer quantum Hall effect" (RIQHE).

The microscopic description and thermodynamics of RIQH states remain topics of great interest\cite{smetbias, csathy,goerbig, Fradkin:1999bl,Spivak:2006kf}.  Most experimental input into these questions has come from monitoring RIQH state collapse at elevated temperature or high current bias\cite{csathy, chickering2013, Cooper:1999ji,Cooper:2003cz,smetbias, esslin}. The temperature-induced transition out of insulating RIQH states is far more abrupt that would be expected for activation of a gapped quantum Hall liquid, consistent with their collective nature. RIQH collapse at elevated temperature is apparently a  melting transition of the electronic system out of the electron solid state\cite{csathy}.

Elevated current biases induce transitions out of the insulating RIQH state that occur via sharp resistance steps, a phenomenon that has been interpreted in terms of sliding dynamics of depinned charge density waves\cite{Reichhardt:2005be}, or alignment of electron liquid crystal domains by the induced Hall electric field\cite{smetbias}. These interpretations assume that bias-induced phase transitions happen homogeneously across the sample. On the other hand, finite currents through a quantum Hall sample generate highly localized Joule heating. Considering the collective nature of RIQH states, this suggests a mechanism for forming inhomogeneous phases across a macroscopic sample.  

Here, we show that resistance signatures of high current breakdown for RIQH states reflect a macroscopic phase separation induced by the bias.  That is, the breakdown process itself is sharply inhomogeneous, with the electronic system after breakdown spatially fractured into regions that are either melted (conducting) or frozen (insulating). For all RIQH states from $\nu=2$ to $\nu=8$, the breakdown propagates clockwise or counterclockwise from the source and drain contacts with a sense that depends on the electron- or hole-like character of the particular RIQH state.  The data are explained by a phase boundary between frozen and melted regions that spreads around the chip following the location of dissipation hotspots.

\begin{figure}[t]
\includegraphics{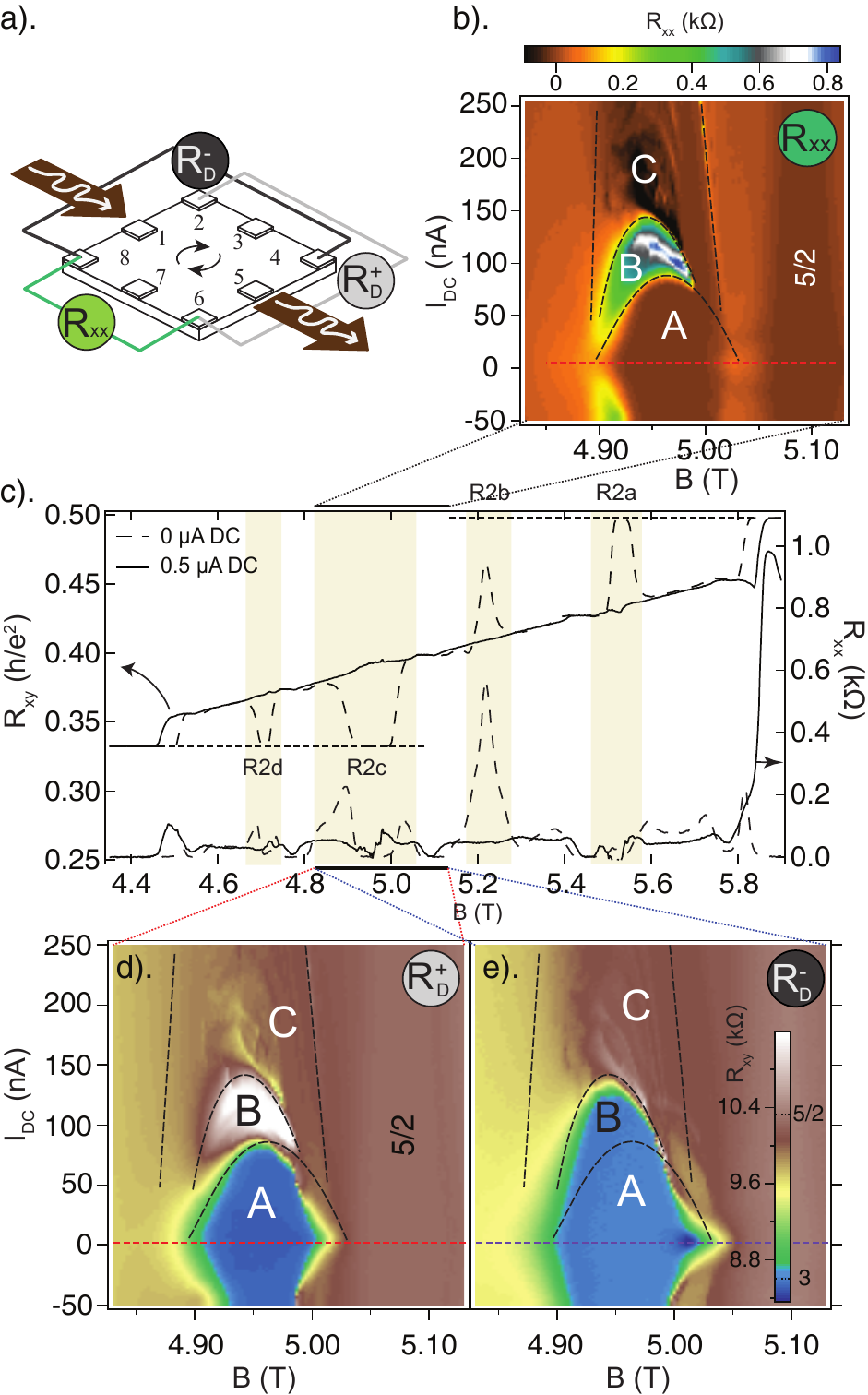}
\caption{a) Measurement schematic combining AC (wiggly arrow) and DC (solid arrow) current bias through contacts 1 and 5.
$\rxx={dV_{86}/dI}$,  $\rdp= dV_{26}/dI$,  and
$\rdm={dV_{84}/dI}$. Curved arrows indicate edge state chirality.  b) Evolution of $\rxx$ with DC bias for the  R2c reentrant and $\nu=5/2$ FQH state, showing  breakdown regions `A', `B', and `C' .  c) $\rxx$ and $\rxy$ ($dV_{37}/dI$) for filling factors $\nu=2-3$, showing the breakdown at high DC bias. (d,e) Simultaneous measurements of $\rdp$ (d) and $\rdm$ (e), taken together with data in panel b). Dashed lines are guides to the eye, denoting identical \{$B,I_{DC}$\} parameters in panels b,d,e.}
\label{4rs}
\end{figure}

Measurements were performed on a 300\,{\AA} symmetrically doped GaAs/AlGaAs quantum well with low temperature electron density $n_s=3.1\times10^{11}$~cm$^{-2}$ and mobility $15\times10^6$~cm$^2$/Vs\cite{manfrarev}. Electrical contact to the 2DEG was achieved by diffusing indium beads into the corners and sides of the 5$\times$5\,mm chip [Fig.~1a]. FQH characteristics were optimized following Ref.~\onlinecite{shiningheating}.    
Differential resistances ${R}\equiv dV/dI_b$ for various contact pairs were measured at 13\,mK by lockin amplifier with an AC current bias, $I_{AC}=5$~nA, at 71 Hz. A DC current bias $I_{DC}$ was added to the AC current in many cases. At zero DC bias, characteristic $\rxx$ and $\rxy$ traces over $2<\nu<3$ show fragile FQH states as well as four RIQH states, labelled R2a-R2d [Fig.~\ref{4rs}c] (refer to the supplement for a complete labelling of  reentrant states). At high current bias the RIQH states disappear, with $\rxy$ moving close to the classical Hall resistance, while most fractional states remain well-resolved.

The RIQH breakdown process can be visualized in 2D resistance maps versus $I_{DC}$ and magnetic field.  Fig.~1 presents several such maps for the hole-like R2c state ($\nu\sim 2.58$), where the Hall resistance reenters to the integer value $R_{xy}=h/3e^2$.  Breakdown transitions for $\rxx$ [Fig.~1b] divide the map into three distinct subregions  [`A', `B', `C'], similar to observations by others\cite{smetbias, esslin}.
Region A is characterized by very low $\rxx$: here the electron solid state is presumably pinned and completely insulating. The sharp transition to region B corresponds to a sudden rise in $R_{xx}$, while for higher bias (region C) the differential resistance drops again to a very small value.

The sharp transitions visible in the RIQH state breakdown [Figs.~1b] are absent from the neighbouring $\nu=5/2$ state.  Considering the range of filling factors investigated here (see supplement), and data from many cooldowns, this behaviour was observed consistently in RIQH states, but never in fractional states, pointing to distinct thermodynamic properties for the two ground states. Qualitative signatures at each pair of voltage probes ($\rxx$, $\rxy$, or the diagonal measurements $\rdp$ or $\rdm$ [Fig.~1a]) did not depend on the specific contacts used in the measurement, but only on the arrangement of the contacts with respect to source/drain current leads (see supplement).

The observation of sharp delineations in the resistance of a macroscopic sample, measured between voltage probes separated by 5\,mm, might seem to imply that the entire sample must suddenly change its electronic state for certain values of bias current and field.  Then one would expect simultaneous jumps in resistance monitored at any pair of voltage probes, albeit by differing amounts.  Comparing the three pairs of voltage probes in Figs.~1b, 1d, and 1e, one sees immediately that this is not the case.   $\rdp$ exhibits  transitions at precisely the same parameter pairs \{$B,I_{DC}$\} as $\rxx$, but for $\rdm$ no resistance change is observed at the dashed line corresponding to the $\rxx$ A-B transition.  It is well known that $\rdp$ and $\rdm$ can be different when the sample is inhomogeneous\cite{beenakkervanhouten, butikker}.  However, the extremely high quality 2DEG samples measured here are  intrinsically homogeneous, as evidenced by the visibility of closely-spaced and fragile fractional states.

\begin{figure}[t]
\includegraphics{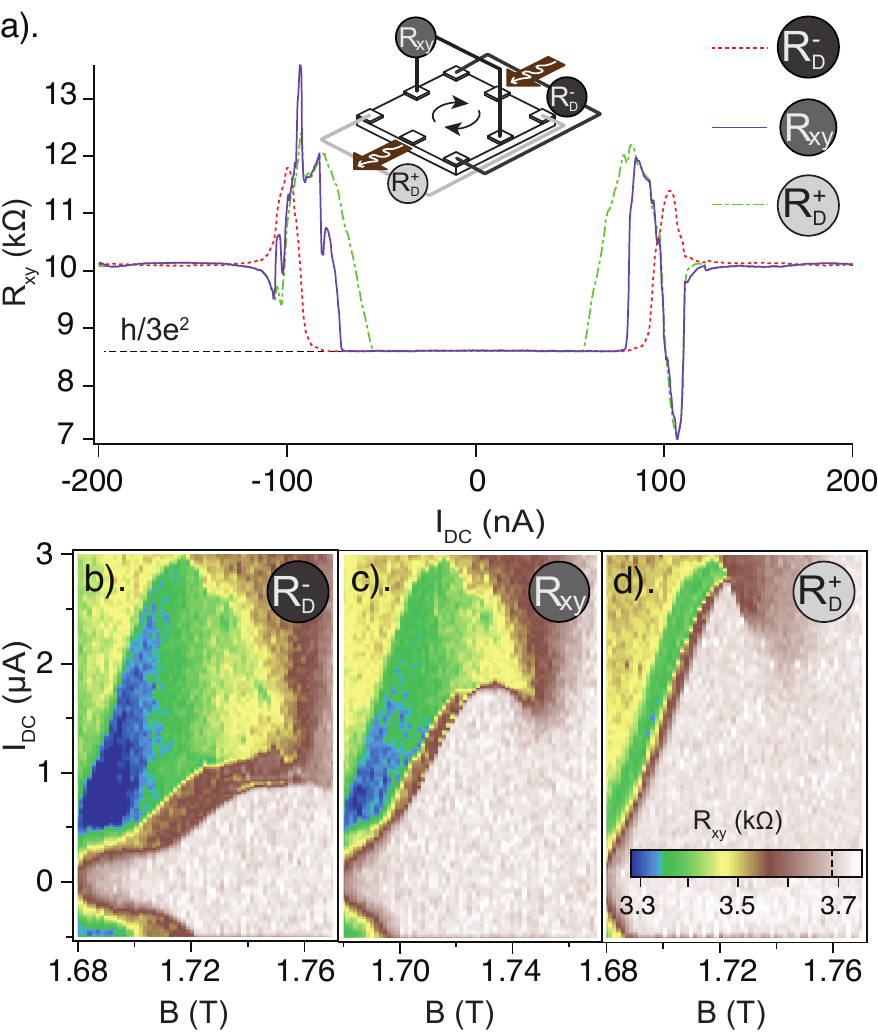}
\caption{(a) Simultaneous measurements showing the evolution of $\rdp$, $\rxy$, and $\rdm$, with DC bias, in the middle of the R2c reentrant state ($I_{AC}$=5\,nA); note that this measurement uses a contact configuration rotated by 90$^\circ$ from Fig.~1. Evolution of (b) $\rdm$, (c) $\rxy$ and (d) $\rdp$ for the R7a reentrant state with DC bias.}
\label{idep}
\end{figure}

$\rdp$ and $\rdm$ contacts are distinguished by the chirality of quantum Hall edge states: moving from source or drain contacts following the edge state chirality, one first comes to the $\rdp$ contacts, then to $\rxy$ contacts in the middle of the sample, and finally to the $\rdm$ contacts. The bias where the A-B transition occurs for $\rdp$, $\rxy$ and $\rdm$ simply follows the spatial distribution of the respective voltage contacts,  as shown in Fig.~2a. An analogous breakdown behaviour (breakdown bias for $\rdp$ lower than for $\rxy$, lower than for $\rdm$) was consistently observed for every hole-like RIQH state. For all electron-like states, a similar breakdown progression was observed but the order was opposite: breakdown bias for $\rdp$ higher than for $\rxy$, higher than for $\rdm$. Figs.~2b-d show this progression for R7a, the electron-like RIQH state between $\nu=7$ and $\nu=8$ [see supplement].

The correlation between electron/hole character and breakdown chirality offers an important hint as to the origin of this effect.  Edge state chirality is fixed by magnetic field direction, and would not suddenly reverse when crossing half-filling for each Landau level.    Instead, we propose an explanation based on localized dissipation in the quantum Hall regime--so-called ``hotspots"--any time a significant bias is applied.

Driving a current, $I_{b}$,  through a sample in the integer quantum Hall (IQH) regime, where $\rho_{xx}$ is close to zero but $R_{xy}$ is large, requires a potential difference $R_{xy}I_{b}$ between source and drain. This potential drops entirely at the source and drain contacts (no voltage drop can occur within the sample since $\rho_{xx}\rightarrow 0$).  Specifically, the voltage drops where the current carried along a few-channel edge state is dumped into the metallic source/drain contact---a region of effectively infinite filling factor.

For a sample in the {\em reentrant} IQH regime, with $\rho_{xx}\rightarrow 0$ as before, hotspots again appear at any location where current flows from a region of higher to lower $R_{xy}$.  But now the local value of $R_{xy}$ is strongly temperature dependent, with a sharp melting transition in both longitudinal and transverse resistances\cite{csathy}.  The electron-like R2a reentrant state, for example, has $R_{xy}^{reentrant}=h/2e^2$ in the low temperature, low bias limit [Fig.~1c], but at higher temperature or bias the state melts to $R_{xy}^{melted}\simeq h/2.35e^2$.  In general, electron-like states have $R_{xy}^{melted}<R_{xy}^{reentrant}$ whereas hole-like states have $R_{xy}^{melted}>R_{xy}^{reentrant}$.  

At low current bias in the RIQH regime, the entire sample is effectively at integer $\nu$ and only the two IQH hotspots are observed, at source and drain contacts.  As the bias increases, the regions around the two IQH hotspots melt and an extra two ``RIQH hotspots" appear where current passes into or out of the melted regions.  This framework, with the 5$\times$5 mm sample broken into macroscopic frozen and melted regions due to dissipation in local hotspots at the boundary, can explain why the breakdown phenomenology was observed only for RIQH states.  A crucial ingredient in this picture is that current flowing from source to drain passes through a well-defined phase boundary, with a large and discrete jump in $\rxy$, even when there are continuous thermal gradients across the sample. This results from sharp electronic phase transitions, which are expected for correlated RIQH states and yield sharp jumps in $\rxx$ and $\rxy$ with elevated temperature, in contrast to activated behaviour of FQH states.

Classical simulations of dissipation in a sample broken into regions with differing $\rxy$ (in this case, $R_{xy}^{reentrant}$ and $R_{xy}^{melted}$) confirm the connection between hotspot location, edge chirality, and changes in filling factor.  Hotspots appear where current passes from melted into frozen regions for the hole-like case in Fig.~3, because $R_{xy}^{melted}$$>$$R_{xy}^{reentrant}$; they appear on the opposite sides of the melted regions for electron-like states where $R_{xy}^{melted}$$<$$R_{xy}^{reentrant}$ (see supplement). The semicircular melted regions in Fig.~3 are defined by the simulation inputs; in reality these regions would be expected to spread in the direction of extra heating, that is, following the hotspot locations, until heat flow into the substrate balances the hotspot dissipation.

\begin{figure}
\center
\includegraphics[scale=1]{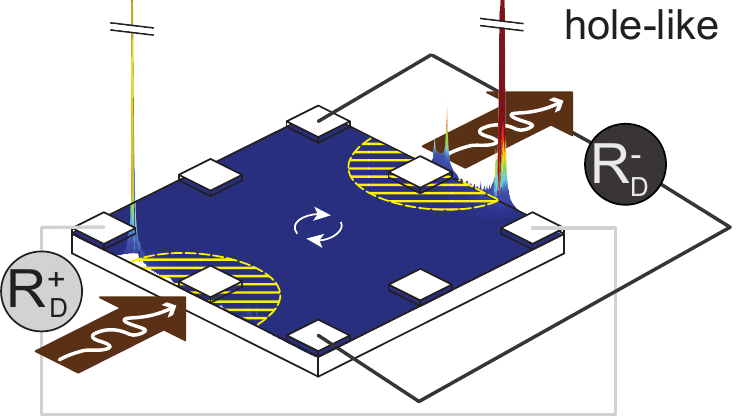}
\caption{Classical simulation of dissipation (colorscale and 3D projection in a.u.) due to current flow in a sample with regions of different $R_{xy}$: hatched semicircles are melted states near each contact with $R_{xy}=h/(2.5e^2)$; the bulk (dark blue) is the (frozen) reentrant state with $R_{xy}=h/(3e^2)$.  Simulation captures hotspot locations but does not accurately capture relative magnitudes of dissipation in different hotspots.}
\label{idep} 
\end{figure}

Consider as an example the R2c measurement in Fig.~1 [Fig.~1c].  RIQH hotspots for hole-like states are downstream from source/drain contacts following edge state chirality [Fig.~3], so the melted/frozen boundaries  propagate clockwise from contacts 1 and 5 around the sample edge.  Within region A, we speculate that the hotspots have not yet passed a voltage probe, so no change is observed in $R_{xx}$, $R_D^+$, or $R_D^-$. When the hotspots pass voltage probes 2 and 6, used for $R_{xx}$ and $R_D^+$, both resistances register a jump due to the potential drop at the hotspot. $R_D^-$ is unaffected, because the potential drop did not pass into or out of the contact pair (4,8).  This mechanism also explains the progression of A$\rightarrow$B transitions for \{$R_D^+,R_{xy},R_D^-$\} in Fig.~2.  For R2c [Fig.~2a], the hotspot first passes the $R_D^+$ probe, then the $R_{xy}$ contact, then the $R_D^-$ contact; for the electron-like R7a [Fig.~2b,c,d], the hotspot propagates against the edge state chirality, so it passes the $R_D^-$ probe, then $R_{xy}$, then $R_D^+$.

Finally, we turn to a measurement configuration that has been used to investigate possible anisotropy in the electron solid at high bias, when the Hall electric field is large.  Ref.~\onlinecite{smetbias} compared $R_{xx}$ measured parallel or perpendicular to a large DC current bias, by rotating the $R_{xx}$ voltage probes and AC current bias contacts by 90$^\circ$ with respect to the DC bias contacts [Fig.~4a,b]. It was observed that the low-$R_{xx}$ region A extended to much higher bias for the ($AC\perp{DC}$) orientation, compared to the conventional ($AC||DC$) orientation.  While Ref.~\onlinecite{smetbias} focused on R4 states exclusively, we found analogous behaviour for all reentrant states measured [see e.g Figs.~4c,d].

This behaviour can be simply explained by the hotspot-movement mechanism outlined above, without resorting to induced anisotropy in the electron solid.  Figs.~4a and 4b schematics include dashed lines to show hypothetical melted-frozen boundaries for an electron-like state at intermediate bias, with associated RIQH hotspots ($\star$).  The melted region surrounds the DC (not AC) current contacts, because the measurement is done in the limit of vanishing AC bias. The boundary is not symmetric around the DC contacts as the melted region is presumed to have propagated counterclockwise (for electron-like states) from the contacts, following the $\star$ hotspot locations.

\begin{figure}
\includegraphics[scale=1]{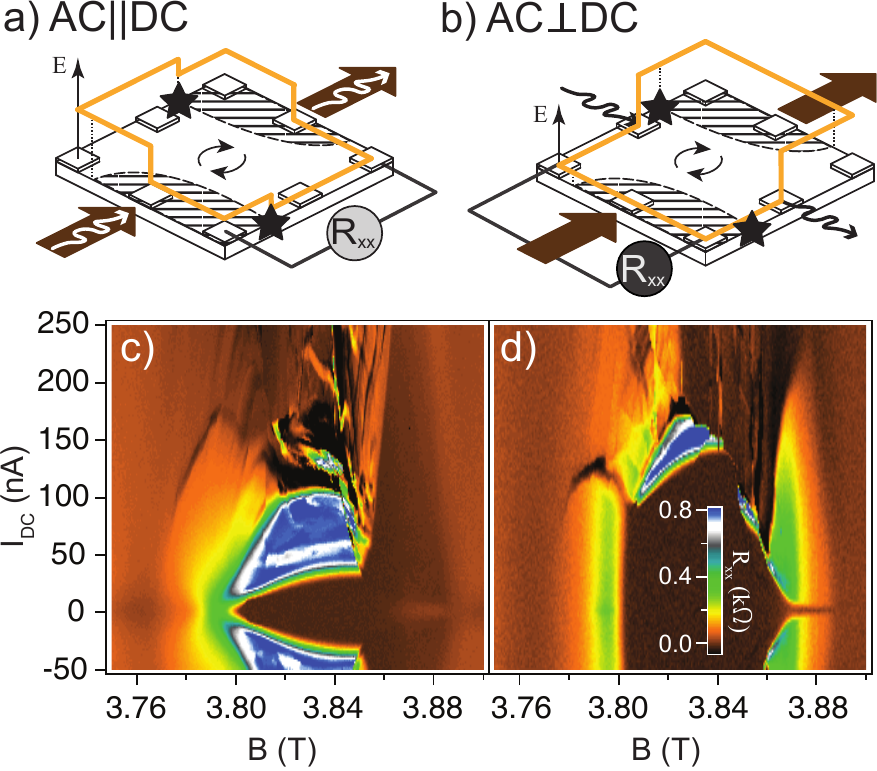}
\caption{Comparison of measurement geometries a) $AC||DC$ and b) $AC\perp{DC}$; arrows label source and drain contacts for DC (solid) and AC (wiggly) bias, and edge state chirality (curved).  Vertical axis `E' denotes local AC edge state potential (yellow line).  Hatched areas are hypothetical melted regions for an electron-like reentrant state at intermediate DC bias, $I_{DC}\sim50$nA in panels (c,d). Hotspots at the melted/frozen boundary indicated by $\star$.  (c,d) R3a $R_{xx}$ maps in $(I_{DC},B)$ plane for c) $AC||DC$ and d) $AC\perp{DC}$ measurement.}
\label{idep} 
\end{figure}

The local AC potential along the edge of the sample drops sharply when passing the AC source and drain (the conventional IQH hotspots), but a second smaller potential drop occurs at each $\star$ when the melted region includes an AC source/drain [e.g.~Fig.~4a].  
For this distribution of melted and frozen phases, there is a potential drop  between the $R_{xx}$ voltage probes in Fig.~4a but not in Fig.~4b, so large $R_{xx}$ would be registered  only when  $AC||DC$. A configuration like that shown in Figs.~4a,b might correspond to intermediate bias, around 50 nA in Figs.~4c,d, thus explaining the large region of high $\rxx$ in Fig.~4c that appears only above 100 nA in Fig.~4d.

In conclusion, we demonstrated that bias-induced breakdown of the RIQH effect is inhomogeneous across mm-scale samples, and propagates chirally from source and drain contacts with a sense that depends on the electron- or hole-like character of the reentrant state.  This phenomenon appears to result from a thermal runaway effect due to phase segregation and dissipation hotspots; it was observed only in (correlated) RIQH states, not fractional or integer states, pointing to their qualitatively distinct thermodynamic properties.   This experiment shows the danger in interpreting macroscopic measurements at a microscopic level, especially where electronic phase transitions are sharp.  On the other hand, it demonstrates the power of using breakdown characteristics as a probe into thermal properties of correlated electron states.  Looking ahead, it would be particularly interesting to investigate combined Corbino/hall bar geometries, where hotspot-induced breakdown could be included or avoided as desired.  Adding multiple small contacts within the interior of a sample would enable melted and frozen phases to be measured separately, and the nature of the transition region to be probed directly. 

\begin{acknowledgments}
The authors acknowledge helpful discussions with J. Smet and A. Stern.  Experiments at UBC were supported by NSERC, CFI, and CIFAR.  The molecular beam epitaxy growth at Purdue is supported by the U.S. Department of Energy, Office of Basic Energy Sciences, Division of Materials Sciences and Engineering under Award DE-SC0006671. 
\end{acknowledgments}


\widetext
\clearpage
\newpage
\setcounter{equation}{0}
\setcounter{figure}{0}
\setcounter{table}{0}
\setcounter{page}{1}
\makeatletter
\renewcommand{\theequation}{S\arabic{equation}}
\renewcommand{\thefigure}{S\arabic{figure}}
\renewcommand{\bibnumfmt}[1]{[S#1]}
\renewcommand{\citenumfont}[1]{S#1}


\setcounter{figure}{0}
\section{Supplement}
\subsection{Reentrant state labelling}
\begin{figure}[h]
\includegraphics[scale=1]{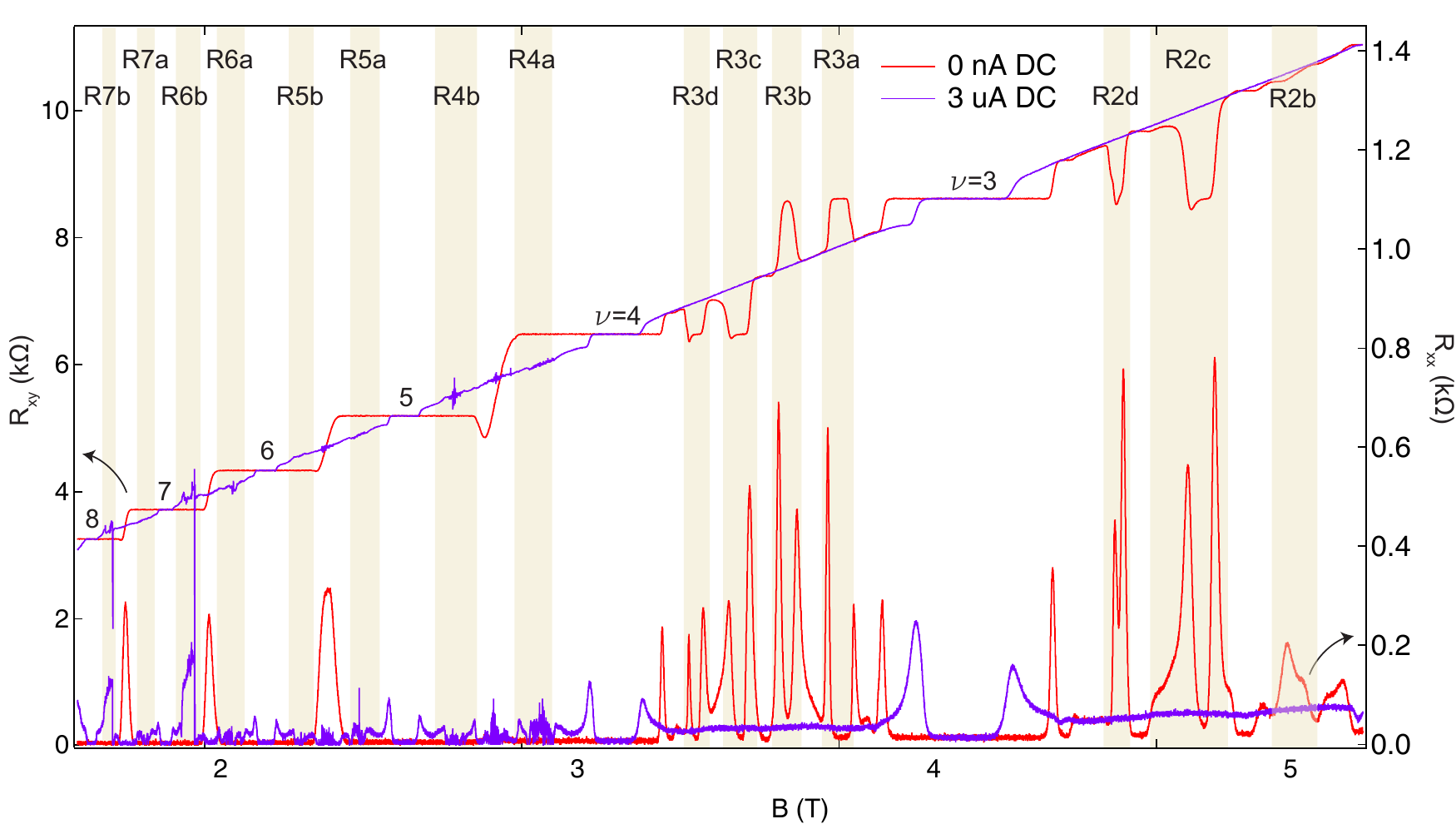}
\caption{$\rxx$ ($dV_{24}/dI$) and $\rxy$ ($dV_{37}/dI$) for filling factors $\nu=8/3-8$.}
\label{s5}
\end{figure}

Each range of magnetic field corresponding to the change of filling factor between two neighbouring integer values has two reentrant states at low  field ($\nu\leq4$) and four reentrant states for each spin branch in the N=1 Landau level, that is, for $4\leq\nu\leq3$ and $3\leq\nu\leq2$ (Fig.~\ref{s5}). Following conventional notation from literature, we label them R2a-R2d and R3a-R3d for the N=1 Landau level and RXa/RXb for reentrant states at higher filling factors where X is the filling factor of the partially-filled Landau level. Figure\,S1 compares zero bias and high bias traces. The shaded lines denote the areas corresponding to each RIQH state.
\subsection{How do $\rdp/\rdm$ measurement depend on contacts?}
Figure S2 compares diagonal measurements of reentrant states between $\nu=3-4$ for contact configurations rotated by 90$^\circ$ around the chip.  Almost all of the major characteristics of the RIQH breakdown are the same for the two sets of contacts, demonstrating that the effects described in the main text do not depend on specific contact imperfections but rather on relative location of voltage and current contacts.

\begin{figure*}[h]
\includegraphics[scale=0.92]{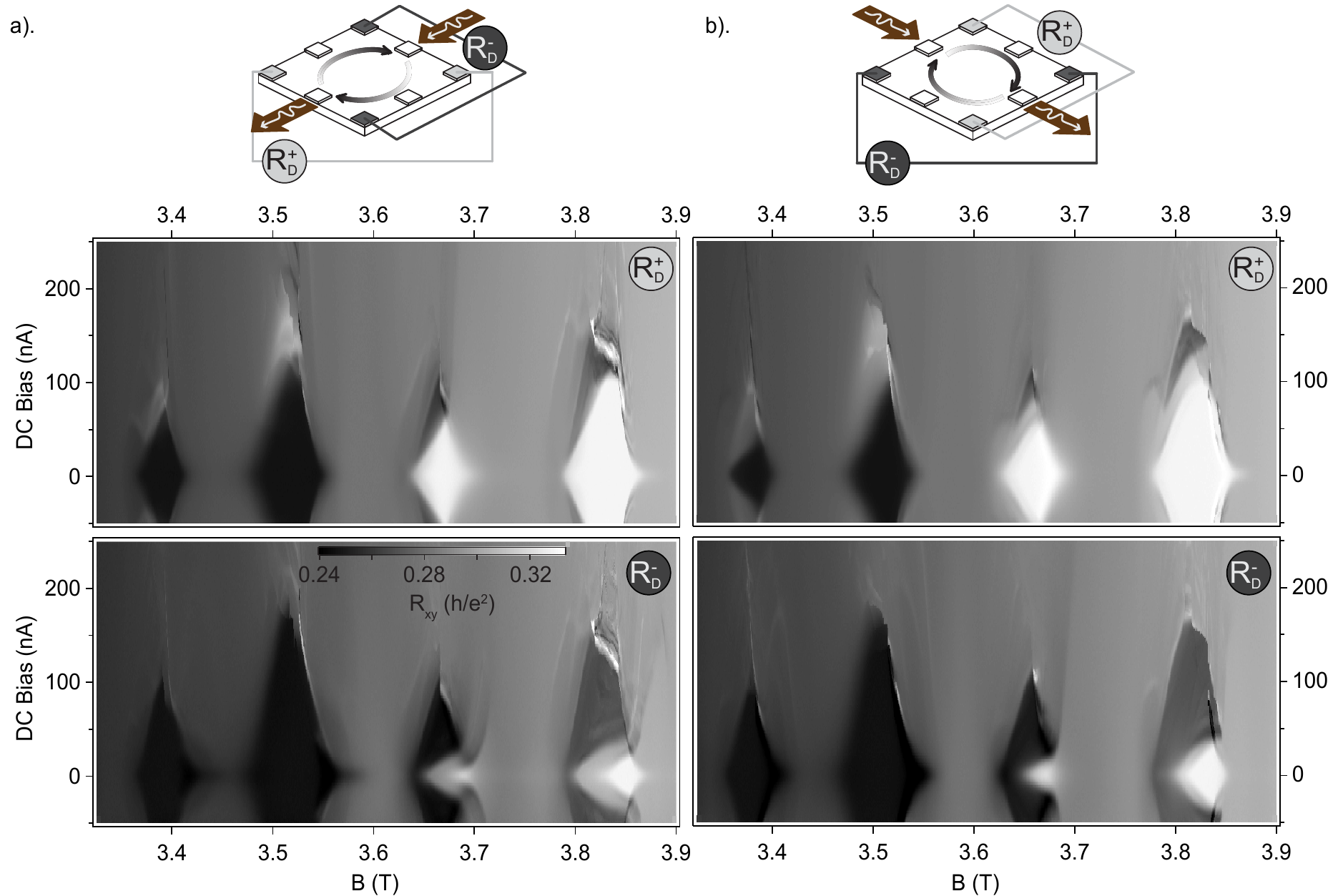}
\caption{Diagonal measurements, $\rdp$ and $\rdm$, with the current flowing in two perpendicular orientations with respect to the sample axes. Note that $\rdp$ and $\rdm$ are, as always, defined with respect to the source and drain contacts.  Most features of the data are reproduced for both orientations, indicating that they depend not on specific contacts but only on the relative location of voltage probes with respect to source and drain.  Specifically, it is clearly seen that which of the two diagonal measurements breaks down at a higher bias (for a given reentrant states) stays the same for the different current orientations.}
\end{figure*}

\subsection{Comparing a broader set of voltage probes}

As further confirmation that a melted pool of free carriers around source/drain contacts spreads chirally around the edge when the bulk of the sample is at a frozen reentrant state, we map out in detail the spreading of potential jumps around the sample for the R2c state.  R2c is the hole-like reentrant state,  corresponding to $\nu=3$ ($R_{xy}^{frozen}=h/3e^2\sim 8.6k\Omega$) in the frozen state, and to $\nu=2.58$ ($R_{xy}^{melted}=h/2.58e^2=10k\Omega$) in the melted state.  Fig.~S3a enumerates sample contacts, with current sourced into contact 1 and drained from contact 5 which is assumed to be at ground potential.  Also shown in Fig.~S3a are three plausible extents of the melted regions, denoted MFM1,2,3 (melted-frozen-melted, MFM), with their associated hotspots where the edge state potential drops.  MFM1,2,3 correspond consecutively to what one might expect for increasing magnitudes of  DC bias current.    As an example, the edge state potential expected for MFM3 is shown in Fig. S3b using arrow line thickness, where $\Delta\equiv I\times(R_{xy}^{melted}-R_{xy}^{frozen})$ comes from the filling factor difference between the melted and frozen regions.

\begin{figure*}[t]
\includegraphics[scale=1]{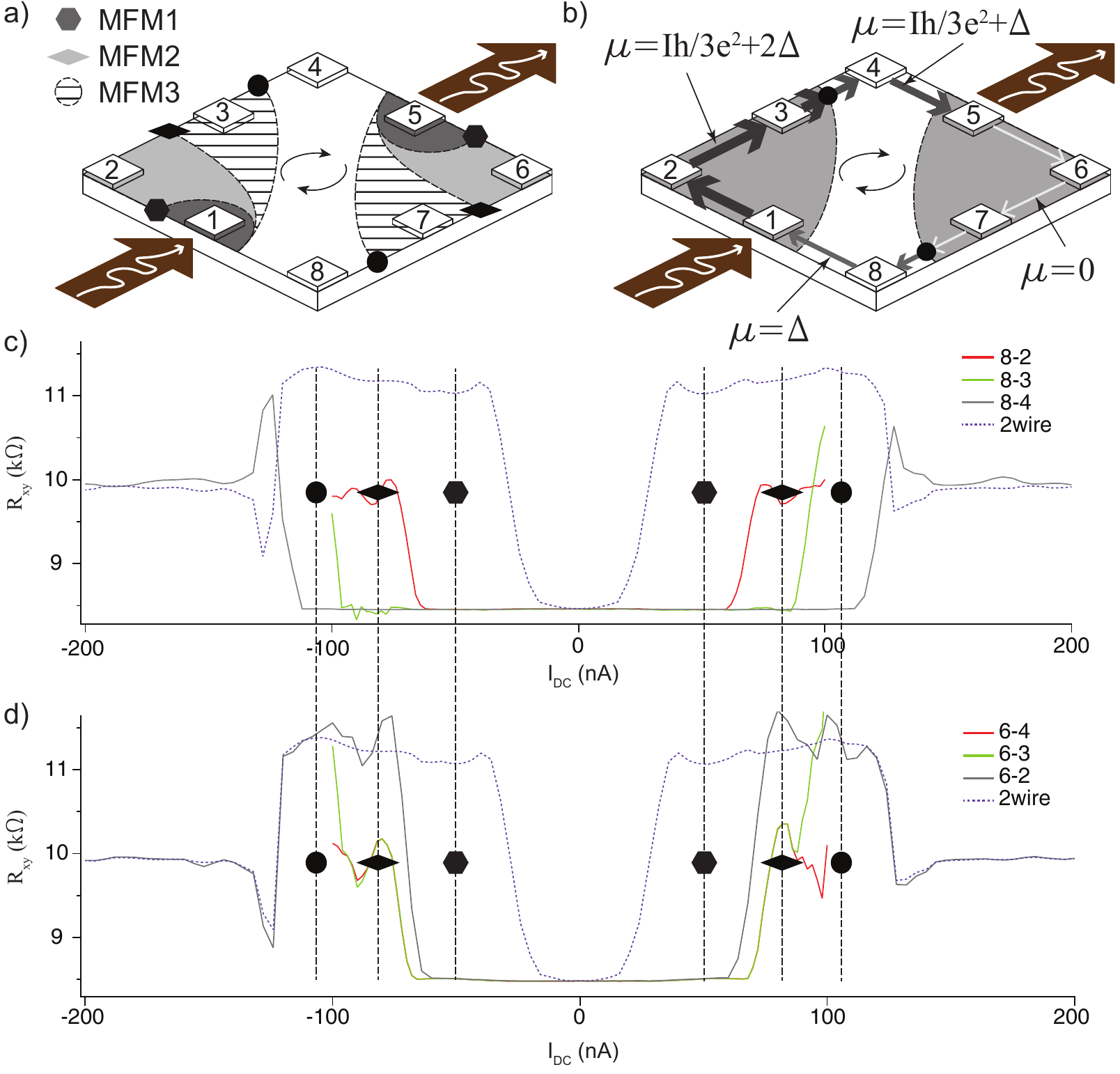}
\caption{a) Breakdown propagation in a hole-like RIQH state; hexagonal, diamond and circle marks depict the position of the hot spots for three different sizes of melted regions (dark grey, light grey and hatched), corresponding to different bias currents. b) Electrical potential distribution along the edges due to bias current $I$, corresponding to MFM3 with hotspot locations marked by black circles. c) and d) present a comparison of breakdown propagation for multiple voltage probes along one side of the sample with respect to a single contact on the opposite side.  c) shows $R_{xy}$ signatures of breakdown for probes 2,3,4 with respect to contact 8; d) shows analogous data with respect to contact 6.  The symbols (circle, diamond, hexagon) are used to denote bias current values that would corresponding to the hot spot locations shown in a).  Two-point measurements (labelled 2wire) corresponding to the voltage across the source-drain contracts are shown by dashed curves in c) and d), see text.}
\end{figure*}

What do the localized points of potential drop along the edge state imply for voltage readings at various contact pairs? For MFM1, corresponding to the dark grey melted regions with the hexagonal hotspots, the potential drops along the upper edge state ``before'' (following edge state chirality) the edge state has reached contacts 2,3,4, and the potential drops along the lower edge state before contacts 6 or 8, so none of the voltage contact pairs 8-2, 8-3, 8-4, 6-2, 6-3, or 6-4 would record a resistance jump relative to the zero bias measurement.  For MFM2, corresponding to the light grey melted region and diamond hotspot, the potential drop in the upper edge state occurs after contact 2 but before contacts 3 or 4; along  the lower edge the potential drop occurs after 6 but before 8.   Considering pairs 8-2, 8-3 and 8-4, we therefore expect a resistance jump only in 8-2, whereas potential jumps have occurred for all three pairs 6-2, 6-3, and 6-4.  The magnitude of the jump in 6-2 is twice what it is in 6-3 or 6-4, because that pair includes jumps along both the upper and lower edge states.  An analogous analysis applies for MFM3.  Here only contact pair 8-4 remains at the original voltage level, while all other pairs record at least one additional potential drop.

Figures S3c and S3d compare the measured breakdown propagation for multiple voltage contact pairs in the R2c state.  Fig.~S2c  records  contact pairs 8-2, 8-3, and 8-4, while Fig.~S2d records 6-2, 6-3, and 6-4.  The hotspot symbols from S2a are also used in S2c,d to suggest a correlation between regions of bias current and hotspot locations for MFM1,2,3 respectively.  For example, we suggest that $I_{DC}=50$nA corresponds to MFM1, with the hexagonal symbol.  At this point, none of the contact pairs have recorded a voltage jump compared to the zero bias case, because the hotspots are ``before" contacts 6 and 2.  $I_{DC}=85$nA corresponds to MFM2, with the diamond symbol; at this point all of 6-2, 6-3, and 6-4 have recorded a voltage jump in S2d whereas only only 8-2 records a voltage jump in S2c. The fact that the voltage jump between the levels seen for hexagonal and diamond markers occurs for bias current around $I_{DC}=70$nA indicates that this is the bias current for which the hotspots pass contacts 2 and 6.  An analogous explanation applies for the circle markers around $I_{DC}=110$nA, which apparently corresponds to MFM3.

Figures S3c and S3d also include a trace (dashed line) for the two-wire resistance of the sample, that is, the potential measured across source and drain contacts 1-5.  A resistance of $2.7k\Omega$ has been subtracted from the trace to take into account wire resistance in the cryostat as well as imperfect ohmic contacts.  At zero DC bias, when the entire sample is in the frozen reentrant state corresponding to $\nu=3$, the two wire resistance must be $R=h/3e^2$ (after subtracting wire and ohmic contact resistances).  But once a finite melted region has spread away from source and drain contacts, resulting in MFM1 with the hexagonal hotspots in Fig. S2a, the two-wire resistance increases to include the two additional potential drops at the hotspots.  Thus the two-wire resistance records the first appearance of a melted region spreading beyond source and drain contacts, around $I_{DC}=25$nA, whereas the voltage probes 2 and 6 only register the movement of the hotspot past these two voltage probes.

\subsection{Simulation details}
The simulation presented in the primary manuscript is intended as an aid to the intuition, for understanding why and where hotspots may form when currents are driven through a sample with inhomogeneous filling factor.  It is not intended to provide a particular insight into quantum (vs classical) Hall phenomenology, but rather to help the reader see the counterintuitive characteristics of local dissipation maps in a situation where the magnetoconductance matrix is dominated by off-diagonal elements.

\begin{figure}[h]
\includegraphics[scale=1]{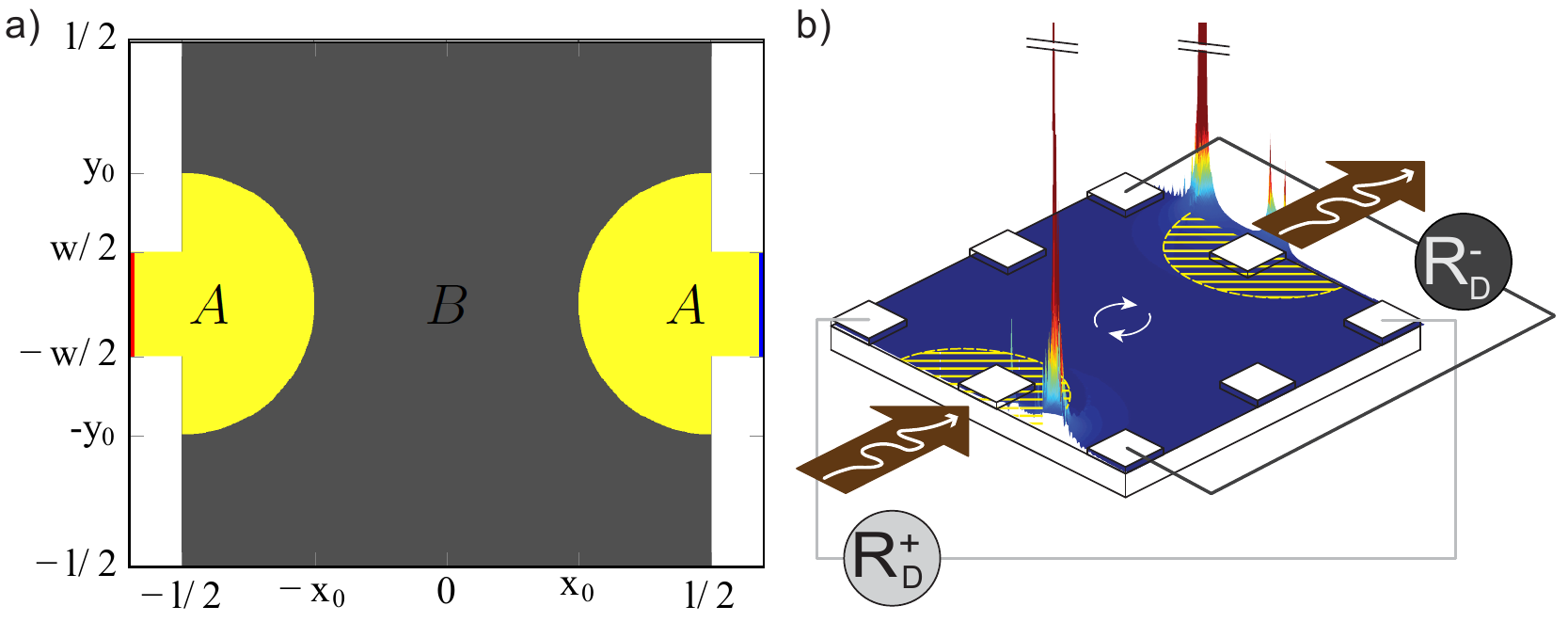}
\caption{a) Geometry of the domain $\Omega$, where simulation is performed. Regions A and B denote the areas with different $\sigma_{xy}$'s in the simulation.  As it relates to the primary discussion in the text, region A would correspond to a melted RIQH state, and region B to the frozen state; b) Classical simulation of dissipation (colorscale and 3D projection in a.u.) due to current flow in a sample divided into regions with different $R_{xy}$: hatched semicircles correspond to melted states near each contact with $R_{xy}=h/(2.5e^2)$, while the bulk (dark blue) is the reentrant state with $R_{xy}=h/(2e^2)$}
\end{figure}

The results are based on a classical solution of transport equations (Kirchhoff's laws). Considering a two dimensional domain $\Omega$, for any point $(x, y)\in\Omega$, Kirchhoff's laws dictate:
\begin{equation}
\nabla\cdot{j}=0, 
\nabla\times{E}=0,
\end{equation}
where j and E are the current density and electric field respectively. Introducing the electric potential, $E=-\nabla\phi$, the equation for E is trivially fulfilled. Finally, assuming the local relation $j=\sigma{E}$, we get:
\begin{equation}
\nabla\cdot{(\sigma\nabla\phi)}=0,
\end{equation}
where $\sigma$ has the general form:
\[
\sigma=
\begin{bmatrix}
\sigma_{xx} & -\sigma_{xy}\\
\sigma_{xy} & \sigma_{xx}
\end{bmatrix}
\]

As long as $\sigma_{xx}>0$, (2) is an elliptic differential equation, and therefore, the existence of an unique solution for $\phi$ is guaranteed for any combination of boundary conditions (BC). Notice that, although we are interested in the quantum Hall effect, at the level of the semi-classical equations we cannot consider the strict limiting case $\sigma_{xx}=0$, since the equation will no longer be elliptic and hence not solvable. As we are interested in high magnetic fields, $\sigma_{xy}/\sigma_{xx}\sim 100$ is used.

Next, the geometry, $\Omega$, and the BC should be defined. The geometry is shown in Fig.~S4a.  In our simulation we consider two cases for BCs:
\begin{enumerate}
\item $\sigma_{xy}^A=2.5e^2/h$, $\sigma_{xy}^B=2e^2/h$
\item $\sigma_{xy}^A=2.5e^2/h$, $\sigma_{xy}^B=3e^2/h$
\end{enumerate}
In both cases $\sigma_{xx}=0.02e^2/h$ for any $(x, y)\in\Omega$. A current I is injected uniformly through the red boundary (contact) and it is collected from blue boundary.

The PDE in Eqn.~(S2) is solved with a finite element method, by casting the PDE into integral forms and optimizing to the weak solution among the finite dimensional vector space of continuous piecewise linear function on a triangular grid.  The outcome of the solution is presented in the main text and Fig.~S4b.

\subsection{RIQHE at higher Landau levels}
Figure S5 shows an example of the evolution of RIQH transport signatures at high Landau levels, specifically between $\nu=8$ (B=1.54\,T, just off the left side of the scan) and $\nu=7$ (B=1.76\,T, just off the right side of the scan).  Above $\nu=4$, there is only one reentrant state on each side of half filling. The two states on either side of half-filling are labeled a,b above $\nu=4$, instead of a,b,c,d for the four RIQH states between $\nu=3$ and $\nu=4$ or $\nu=2$ and $\nu=3$.

At low biases reentrant states are indistinguishable from, and directly connected to, integer plateaus, with quantized transverse and vanishing longitudinal resistance. However, when the bias current is increased the pattern of RIQH state breakdown is similar to that observed for melting at elevated temperatures: the region of the RIQHE stability shrinks in the magnetic field and reentrant states become separated from integer plateaus.

\begin{figure}[h]
\includegraphics[scale=1]{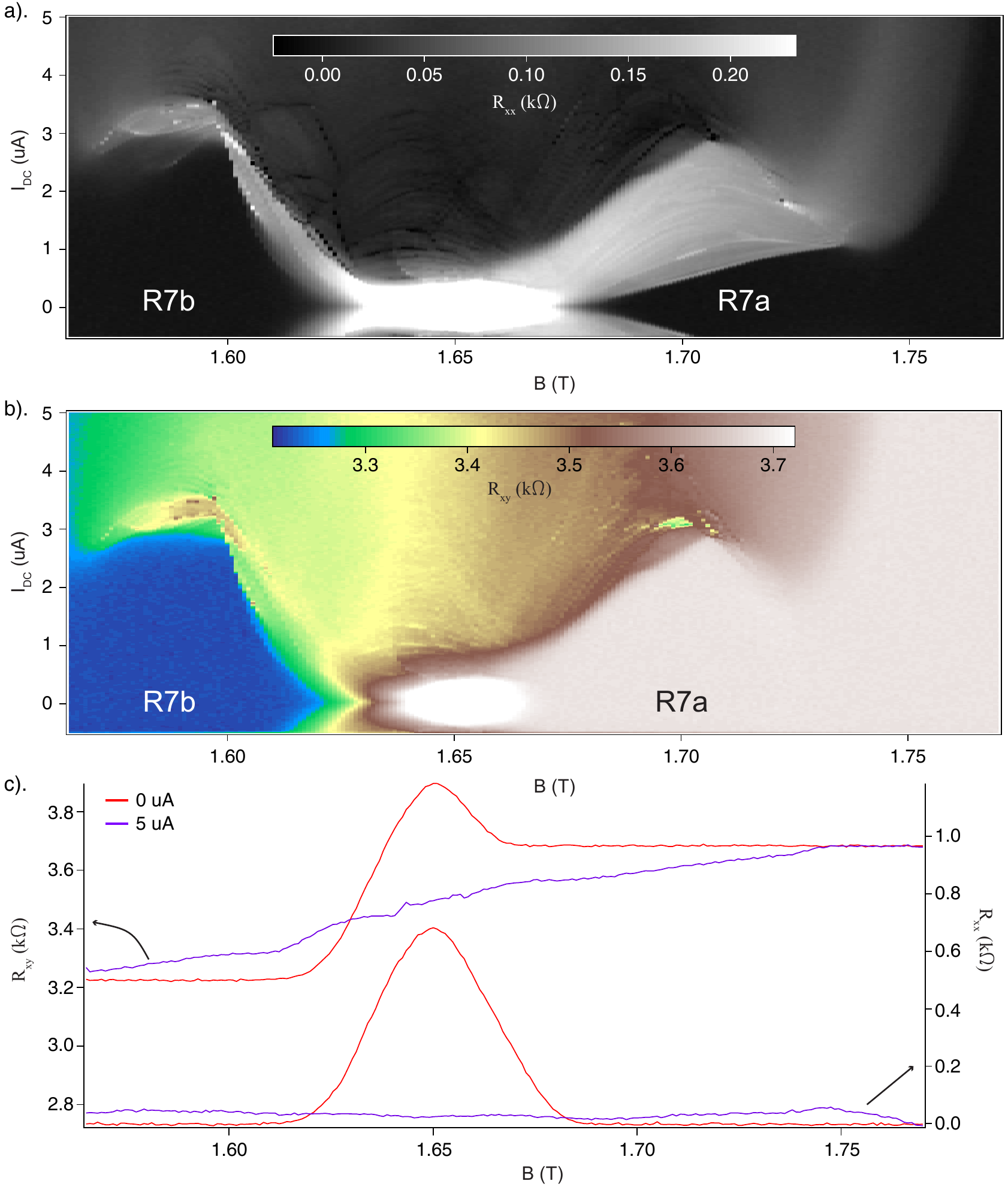}
\caption{$\rxx$ ($dV_{24}/dI$) and $\rxy$ ($dV_{37}/dI$) for filling factors $\nu=7-8$, showing the breakdown at high DC biases: a) $\rxx$, b)$\rxy$ and c) cross-sections of $\rxx$ and $\rxy$ at 0 and 5 $\mu$A DC bias.}
\end{figure}

\end{document}